\documentclass{article}
\usepackage{mrs2005,epsfig}

\def\ergs{ergs s$^{-1}$}

\def\msun{\ifmmode M_{\odot} \else M$_{\odot}$\fi}
\def\zsun{\ifmmode Z_{\odot} \else Z$_{\odot}$\fi}
\def\lsun{\ifmmode L_{\odot} \else L$_{\odot}$\fi}
\def\lya{\ifmmode {\rm Ly}\alpha \else Ly$\alpha$\fi}

\begin{document} 
\title{Looking for z$>$7 galaxies with the Gravitational Telescope}

\author{R. Pell\'o$^{1}$, J. Richard$^{1}$, D. Schaerer$^{2,1}$, J. F. Le
  Borgne$^{1}$, J. P. Kneib$^{3}$, A. Hempel$^{2}$ }
\affil{(1) Observatoire Midi-Pyr\'en\'ees, Laboratoire
d'Astrophysique, UMR 5572, 14 Avenue E. Belin, F-31400 Toulouse, France} 
\affil{(2) Geneva Observatory, 51 Ch. des Maillettes, CH--1290 Sauverny, Switzerland}
\affil{(3) OAMP, Laboratoire d'Astrophysique de Marseille, UMR 6110 traverse
du Siphon, 13012 Marseille, France}  
 
\begin{abstract} 
We summarize the main results obtained recently by our group 
on the identification and study of very high-$z$ galaxies (z$>$7) using lensing
clusters as natural gravitational telescopes. A description of our pilot 
survey with ISAAC/VLT is presented, aimed at the spectroscopic
confirmation of z$>$7 candidate galaxies photometrically selected from deep
near-IR, HST and optical ground-based imaging. The first
results issued from this survey are discussed, in particular the global
photometric properties of our high-$z$ candidates, and the implications for
the global star formation rate at very high-$z$. 

\end{abstract} 
 
\section{Introduction} 
 
   Considerable advances have been made during the last decade in the
exploration of the early Universe, from the discovery and detailed
studies of redshift $z \sim 3$ galaxies (the so-called Lyman break galaxies,
LBGs, e.g. Steidel et al.\ 2003),
over $z \sim$ 4--5 galaxies found from numerous deep multi-wavelength
surveys, to galaxies at $z \sim$ 6--7, close to the end of
reionisation epoch of the Universe (e.g. Hu et al.\ 2002, Kodaira et al.\ 2003,
Cuby et al.\ 2003, Kneib et al. 2004, Stanway et al.\ 2004, Bouwens et al.\
2004b). Extending the searches beyond $z\simeq$ 6.5 and back to ages where the
Universe was being re-ionized (cf.\ Fan et al.\ 2002) requires extremely deep
observations in the near-IR bands. Indeed, astounding depths can be reached in
ultra-deep fields, such as demonstrated e.g.\ recently with J and H imaging of
the NICMOS Ultra-Deep Field (UDF; Thompson et al.\ 2005; Bouwens et al. 2004a,
Bouwens et al. 2005) from which 5 faint ($H_{AB} \sim$ 27) candidates at $z
\sim$ 7--8 have been identified (Bouwens et al.\ 2004b). 

We present in this paper a summary of our results on a deep survey of lensing
clusters with ISAAC/VLT, aimed at constraining the abundance of star-forming
galaxies at $z\sim6-11$ taking benefit from lensing magnification to improve
the search efficiency and subsequent spectroscopic studies (see more details
in Pello et al. 04 and Richard et al. 05). We briefly describe the photometric
technique used to identify very high-$z$ objects, the construction and analysis of the
photometric catalogs, the luminosity funtions (LFs) derived for star-forming
galaxies up to $z\sim10$, and the implications for the cosmic SFR. Throughout
this paper we adopt standard cosmological parameters ($\Omega_{\Lambda}=0.7$,
$\Omega_{m}=0.3$, $H_{0}=70\ km\ s^{-1}\ Mpc^{-1}$).

\section{Photometric Survey and Selection of High-z Candidates} 

   The objective of our Survey was to obtain deep near-IR photometry from 1.0
to 2.4 $\mu$m, in order to derive accurate photometric redshifts for optical
dropouts in the critical domain $6 \le z \le 11$. 
Simulations have been done to define the observing strategy to target high-$z$
sources using the evolutionary synthesis models
by Schaerer (2002, 2003) for Population III and extremely metal deficient
starbursts, together with the usual templates for normal galaxies. 
The main relevant signatures of genuine star-forming sources at $z>7$, which 
are common to all models, are well known: they are optical dropouts,
displaying a strong break and 
``red'' optical vs. IR colors, whereas they exhibit a ``blue'' SED redwards
due to UV rest-frame emission. Different redshift intervals are defined
using the appropriate set of near-IR filters in combination with optical data
through the Lyman break technique, by constraining both the position of the
break and the restframe UV slope.
Figure 1 illustrates this technique in the $8 \le z \le 11$ domain. 

The two lensing clusters used in this study were 
AC114 ($z=0.312$) and Abell 1835 ($z=0.252$). AC114 is a
well-known gravitational telescope, with a lens model well-constrained by
a large number of multiple-images at high-$z$ (Smail et al.\
1995, Natarajan et al.\ 1998, Campusano et al.\ 2001).
A1835 is the most X-ray luminous cluster in the $XBACS$ sample
(Ebeling et al.\ 1998), thus potentially one of the most efficient gravitational
telescopes. We observed these clusters in the
$\sim$ 0.9 to 2.2 $\mu$m domain between
September 2002 and April 2004, covering as far as possible the $z$, $SZ$, $J$,
$H$, and $K$ bands. Optical images 
between $U$ and $I$ bands were available from previous surveys and
data archives (Table 1).

\begin{table}[ht]
{\small
\caption{\label{images}Summary of the photometric dataset used in this
Survey: filter, exposure time, seeing (original images), pixel size,
1 $\sigma$ limiting magnitude within 1.5 \arcsec \ diameter aperture, filter effective wavelength,
AB correction ($m_{AB}=m_{Vega}+C_{AB}$), and references (see Richard et
al. 2005 for details). }

\begin{tabular}{llrllcrrl}\hline
 & Filter & $t_{exp}$ & seeing & pix & depth &  $\lambda_{eff}$& $C_{AB}$ &  Reference \\
 & & [ksec] & [\arcsec] & [\arcsec] & [mag] & [nm] & [mag] & \\
\hline
\hline
AC114& $U$ & 20.00 & 1.3 & 0.36 & 29.1 & 365 & 0.693 &  Barger et al.\ 1996\\
 & $B$& 9.00 & 1.2 & 0.39 & 29.0 & 443 &-0.064 &  Couch et al.\ 2001\\
 & $V$& 21.60 & 1.1 & 0.47 & 28.5 & 547 & 0.022 &  Smail et al.\ 1991\\
 & $R_{702}^{2}$ & $\geq$ 24.90 & 0.13 & 0.100 & $\geq$ 28.4 & 700 & 0.299 &  Natarajan et al.\ 1998\\
 & $I_{814}$ & 20.70 & 0.3 & 0.100 & 26.8 & 801 & 0.439 &  Smail et al.\ 1991\\
 & $J$ & 6.48 & 0.52 & 0.148 & 25.5 & 1259 & 0.945 &  Richard et al. 05\\
 & $H$ & 13.86 & 0.40 & 0.148 & 24.7 & 1656 & 1.412 &  Richard et al. 05\\
 & $Ks$ & 18.99 & 0.34 & 0.148 & 24.3 & 2167 & 1.873 &  Richard et al. 05\\
\hline
\hline
A1835& $V$ & 3.75 & 0.76 & 0.206 & 28.1 & 543& 0.018 &  Czoske et al.\ 2002\\
 & $R$ & 5.40 & 0.69 & 0.206 & 27.8 & 664& 0.246 &  Czoske et al.\ 2002\\
 & $R_{702}$ & 7.50 & 0.12 & 0.100 & 27.7 & 700& 0.299 &  Smith et al.\ 2003\\
 & $I$ & 4.50 & 0.78 & 0.206 & 26.7 & 817& 0.462 &  Czoske et al.\ 2002\\
 & $z$ & 6.36 & 0.70 & 0.252 & 26.7 & 919& 0.554 & Richard et al. 05\\
 & $SZ$ & 21.96 & 0.54 & 0.148 & 26.9 & 1063& 0.691 &  Richard et al. 05\\
 & $J$ & 6.48 & 0.65 & 0.148 & 25.6 & 1259& 0.945 &  Richard et al. 05\\
 & $H$ & 13.86 & 0.50 & 0.148 & 24.7 & 1656& 1.412 &  Richard et al. 05\\
 & $Ks$ & 18.99 & 0.38 & 0.148 & 24.7 & 2167& 1.873 &  Richard et al. 05\\
\hline
\end{tabular}
}
\end{table}

Near infrared photometry of extremely faint sources requires a careful data
reduction, described in details by Richard et al. (2005). In summary, after
a standard pre-reduction (ghost, dark and flat-field corrections),
we used the IRAF package XDIMSUM\footnote
{XDIMSUM is a modified version by the IRAF group of the Deep
Infrared Mosaicing Software package by P. Eisenhardt et al. See
ftp://iraf.noao.edu/extern-v212/xdimsum for details} for a
two-step sky-subtraction. During the first pass, each image is
sky-subtracted using the sky pattern obtained from a group of adjacent
frames and a bad-pixel mask is created in the process.
Images are registered and combined
using integer shifts values to preserve the noise properties, with
bad-pixel rejection. Then, sources are 
detected in order to create an object mask, and a second
sky-subtraction is applied to the data. Several versions of
the final images were produced, using slightly different reduction 
recipes, in order to cross-check the final catalogs. 

Photometry in the near-IR bands was obtained after
matching all images to a common seeing with a gaussian
convolution, the worst case being the $J$ band for both clusters.
The $SExtractor$ package (Bertin \& Arnouts 1996) was used for
source detection and photometry. We optimized the
parameters to detect very faint unresolved sources, in order to build an
$H$-band selected sample. The original images were
used to derive the error bars in each band through detailed simulations.
Also limiting magnitudes in the Table~1 and completeness levels in the
different filters were obtained in this way.
The final catalogs include photometry within 1.5\arcsec\ apertures for
all objects detected in the $H$ band; we were able to measure photometry of very faint
sources ($J \sim 24.4-24.8$, $H$ and $Ks$ $\sim$ 23.5) with a
relatively good accuracy (S/N$\ge$ 3-4).
The fraction of spurious detections expected in our photometric catalogs, for
objects detected {\it only} in the reference filter $H$, was estimated from a
special $H$ band \textit{noise image} where all astronomical sources were
removed by subtracting by pairs sequential images acquired with similar seeing
conditions, and then coadding them using the same procedure as for
astronomical images. The result is an image with
the same noise properties compared to the final stacks, but without
astronomical sources. We find that for the faintest magnitude bins considered in
this survey, the \textit{maximum number of false-positive detections} should be
typically lower than 50 \%, for sources detected {\it only} in the reference
filter $H$, and no spurious detections are expected up to
$H=23.0$ (AB$\sim 24.4$). 

A catalog of optical dropouts was selected (i.e., objects 
non-detected in all the optical images, from $U$ to $z$ bands). 
First-category sources are those detected in at least two near-IR bands. Among them,
$\sim$ 89\% of sources detected in the $H$-band reference image in A1835 (and
75\% in AC114) are also re-detected in the pseudo-$\chi^2$ 
image. The later was obtained from the combination of individual $J$, $H$ and
$Ks$ images, normalized by the noise 1$\sigma$ image, and weighted by the
square root of the corresponding exposure-time maps. 
After careful manual inspection, the final catalog of first/second-category
dropouts for Abell 1835 and AC114 contains 18/6 and 8/2 sources respectively,
and number of third category sources (dubious after manual
inspection, detected only in the reference filter) 
close to the critical lines. 

The position of optical dropouts in the different color-color diagrams
provides an estimate of their photometric redshift, and an objective
criterium to classify them into different $z$ intervals.
The majority of optical dropouts in the two clusters fulfill the high-$z$ requirements in
Fig. 1. Those located in the $0\le z \le 8$ region of the diagram
fulfill the ERO selection criterium ($R-K>5.6$), and some of them could be
intermediate-redshift dusty starbursts. The $SZJH$ and $zSZJ$ color-color diagrams were
used to select candidates in the range $7\le z \le 8.5$ and 
$6\le z \le 7.5$ respectively. For about 30$\%$ of our candidates,
the S/N is enough to derive photometric redshifts based on 
SED-fitting using an adapted version of the public 
software $Hyperz$ (Bolzonella et al.\ 2000). 
Objects unambiguously identified as low-$z$ galaxiess are excluded from
the sample, as well as sources whose nature
could not be determined with the present data (either variable sources or
``bright'' EROs with ambiguous but plausible low-$z$ solutions). 

High-$z$ candidates were selected based {\it only} on their photometric
properties, irrespective of their positions with respect to the critical
lines. However, objects located
close to the high-$z$ critical lines are of greater interest, because
of the larger magnification.
The {\it minimum} magnification factor over the region covered by our near-IR
survey is at least $\sim$ 0.7 magnitudes, and at least $\sim$ 1 magnitude over
50 \% of the ISAAC field of view. Thus, the {\it effective} $3 \sigma$ limiting
magnitudes reached here are, at worst, similar to those attained in the HDFS
(Labb\'e et al. 2003) in $JHKs$ (respectively AB$\sim$ 26.8, 26.2 and 26.2).
Our $3 \sigma$ limiting magnitudes in the $H$ band are also very close to the
typical magnitudes of the $z\sim7-8$ $z$-dropouts detected by Bouwens et al.\
(2004b) in the Hubble Ultra Deep Field, with $H_{160}(AB) \sim$ 26.0 to 27.3,
after correction for a typical magnification factor of at least $\sim$ 1
magnitude. 

%
%
\begin{figure}  
\begin{center}
\epsfig{figure=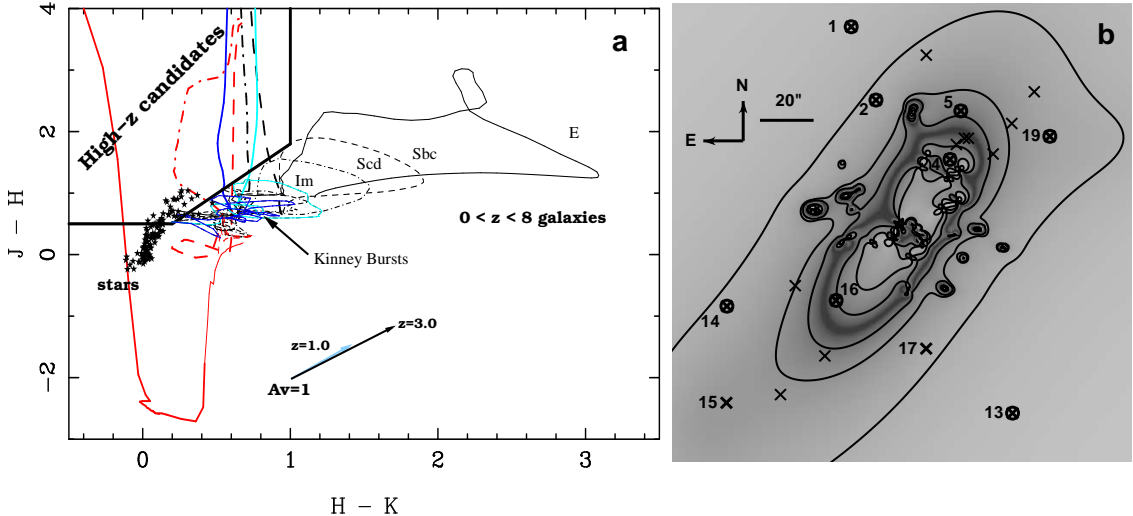,width=15.0cm}  
\end{center}
\vspace*{0.25cm}  
\caption{{\bf a:} 
$J-H$ versus $H-Ks$ color-color diagram (Vega system) showing the position
expected for different objects over the interval $z \sim$ 0 to 11. 
The position of stars and normal galaxies up to $z\ \le$ 8 are shown, 
as well as the shift direction induced by $A_V=1$ magnitude extinction. Thin
and thick lines display models below and above $z=8$ respectively.
Several models for Pop III starbursts are presented, for different fractions
of Lyman-$\alpha$ emission flux entering the integration aperture: 
100\% (red solid line), 50\% (red dashed line) and 0\% (red dot-dashed line).
The location of Kinney et al.\ (1996) starbursts templates is also given for
comparison (SB1(cyan) and SB2 (blue)). All star-forming models enter the 
high-$z$ candidate region at $z\ge8$.
{\bf b:} 
Location of first and second category photometric candidates in the lensing
cluster AC114 (circles and crosses respectively). Contours are overplotted for
magnification values of 1, 2 and 3 magnitudes, computed assuming sources at $z=9$,
although the position of these lines is weakly sensitive to source
redshift within the relevant $z \sim 6-10$.
} 
\end{figure} 

\section{Luminosity Function and Cosmic Star Formation Rate}

The typical magnification values of our candidates range between 1.5 ($\sim$
0.44 mags) and 10 (2.5 mags). For some objects very close to the critical
lines, we found magnifications values $\mu>25$. 
Interestingly, although the selection criteria are only based on near-IR colors
irrespective of magnitudes, {\it almost all} the photometric candidates fulfilling
our selection criteria turn out to be {\it fainter} than $H = 23.0$ (AB $\sim
24.5$). Only three exceptions are found in Abell 1835 among the possible
low-$z$ EROs, as described above. After correction for magnification accross
these fields, the lack of ``bright'' sources means that we have not detected
young starbursts at $z \sim6-10$ more massive than typically a few 10$^8$
M$_{\odot}$ (under standard assumptions for the IMF).

We derived the unlensed $L_{1500}$ luminosity, at 1500 \AA\ restframe, for
all high-$z$ candidates, using the adopted photometric redshift. $L_{1500}$
luminosities were converted into Star Formation Rate (SFR) through the 
usual calibration from Kennicut (1998). The typical SFR obtained for
objects included in the final sample is $\sim$ 10 M$_{\odot}\ $yr$^{-1}$, with
extreme values ranging from a few units to $\sim$ 20 M$_{\odot}$ yr$^{-1}$.
The restframe UV slope of our candidates is extremely blue, usually ranging
between $-1.5$ and $-3.5$, a systematic trend also reported by Bouwens et al.\
(2004b) for their sample of $z\sim7-8$ candidates. 
Although optical-dropouts are stretched by the magnification
factor $\mu$, they appear as point-like sources
in our ground-based images. The physical size of these objects at $z>7$ is
likely to be smaller than 1.7 kpc, with the magnification
factors involved.

Magnification and dilution effects by the lensing field were carefully taken
into account to compute number densities and derived quantities, in particular
to estimate the LF at 1500 \AA\ . The observed sample of candidates was also 
corrected for incompleteness using mock simulations (see details in
Richard et al. 2005).  The combined $L_{1500}$ LFs for both clusters, with the
corresponding error-bars, are given in Fig.~2. 
Only first-priority candidates have been considered, but the difference
obtained when using the full sample is within $1\sigma$ error bars.
STY fits (Sandage, Tammann \& Yahil\ 1979) to the data are also presented in
Fig.~2. The typical value 
found for $L^{*}$ is $10^{41.5}$ \ergs\ s$^{-1}$\ \AA\ $^{-1}$, with a fixed
value $\alpha=1.6$ (i.e., Steidel et al.\ (1999) determination for LBGs at
$z\sim 4$). The STY fit to the data is in remarkably good agreement with the
LF found by Steidel et al. for LBGs at $z\sim 4$, with the usual $(1+z)^{-1}$
correction to account for the surface brightness increase with redshift due to
size scaling for a fixed luminosity, {\it without any additional
renormalization}. A fairly good agreement is also found when comparing with the
LF derived by Bunker et al. (2004) for their sample of $z\sim 6$ candidates in
the UDF, i.e. a density of sources $\sim 1/6$ (-0.8 dex) smaller as compared
to LBGs at $z\sim 4$, still within our $1\sigma$ error bars.
We compare the observed LFs to the predictions obtained from a 
simple model for halo formation based on Press-Schechter formalism, 
assuming that all haloes convert a constant fraction $\sim 0.1$ of their
baryonic mass into stars somewhen between $z=17$ and $z=6$, with a correction for
the visibility time of starbursts, and using two extreme IMF assumptions: a
standard Salpeter and a top-heavy IMF (stars between 50 and 500 M$_\odot$). 
``Top heavy'' IMF models provide a better fit for the bright end of the LF, but
this simple model can hardly explain simultaneously the behaviour of
the bright and faint ends of the LF. 

   Figure2 displays the upper limits for the Cosmic SFR value obtained for each
redshift bin, by integrating the LFs down to $0.3\ L^{*}_{z=3}$, compared to
other surveys.  

%
%
\begin{figure}  
\begin{center}
\epsfig{figure=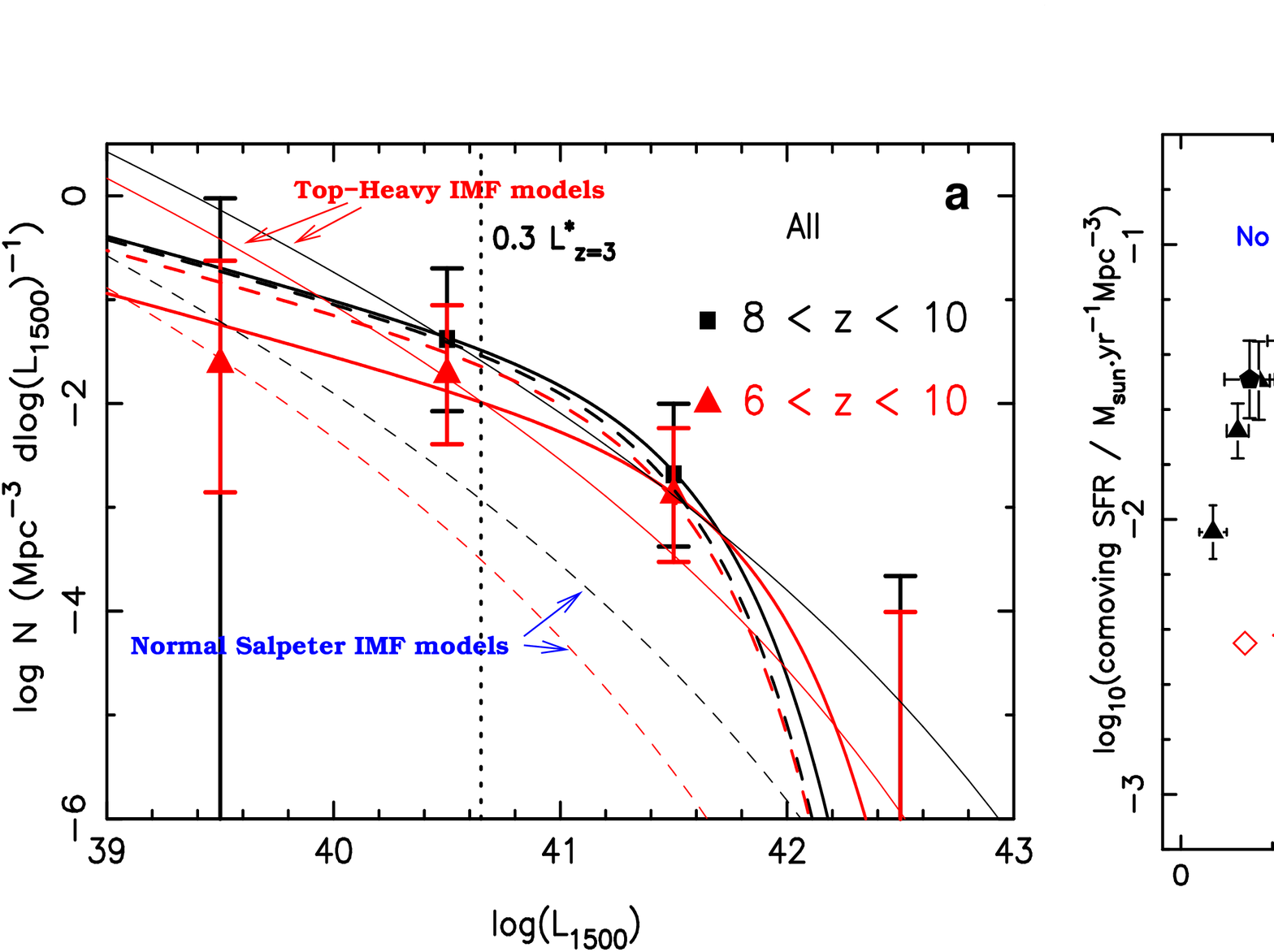,width=15.0cm} 
\end{center}
\vspace*{0.25cm}  
\caption{ 
{\bf a:} $L_{1500}$ LFs derived for the photometric sample of high-$z$
candidates in this survey (adapted from
Richard et al. 05). $z$ intervals shown are: $z=6-10$ (red) and $z=8-10$
(black). Errors include Poisson noise statistics and lensing
uncertainties. STY fits to the LF data are presented by thick solid lines,
and compared to the LF by Steidel et al. for LBGs at $z\sim 4$ (thick dashed
line), with the usual $(1+z)^{-1}$ correction, {\it without any additional
renormalization}. We also display the LFs corresponding to simple models, for
2 extreme IMF assumptions discussed in the text.
{\bf b:} Evolution of the comoving SFR density as a function
of redshift. Our upper limits are compared to other surveys, uncorrected for
extinction (adapted from Bunker et al.\ 2004, and references herein): data
compiled from the CFRS (Lilly et al.\ 1996, filled triangles), Connolly et
al.\ 1997 (filled pentagons), LBG work from Steidel et al.\ 1999 (open
squares), Fontana et al.\ 2003 (open circles), Iwata et al.\ 2003 (cross),
Bouwens et al.\ 2003a (filled diamonds), GOODS (Giavalisco et al.\ 2004, open
stars), ACS estimates by Bouwens et al.\ 2003b (filled stars) and
Bunker et al.\ 2004 (filled circle). The value derived by Bouwens et
al. (2004b) in the UDF is also shown.
Our results are displayed for all (solid) and first-category (dashed)
candidates.
} 
\end{figure} 

\section{Discussion and Conclusions}

Taken at face value, the cosmic SFR density found in this survey is in good agreement
with the theoretical estimates for the redshift domain considered here derived
by Barkana \& Loeb (2001; see their Fig. 29), for a reionization
redshift $\sim 8-6$. However, there is a discrepancy by a factor of $\sim 10$
between our results and previous studies at similar redshifts, in particular
in the UDF (Bouwens et al. 2004, 2005). In all cases, the sources detected are
photometric candidates, and thus upper limits to the actual UV flux densities. The
effective fields surveyed are dramatically small, thus leading to strong field-to-field
variations in the number of sources. Cluster-to-cluster
fluctuations are clearly seen in our sample, although lensing and photometric
considerations could account for most of them. A positive magnification
bias could still be present in this survey (the incompleteness of our sample
in the relevant magnitude domain is smaller than in blank-field surveys),
producing a systematic trend as compared to blank fields.  
Could this result be confirmed on a larger sample of lensing
clusters and blank fields, the slope of the number counts at $z\sim 6-10$
could be precisely constrained, at least for the brightest part of the LF.

Up to now, our spectroscopic survey with ISAAC has targeted 2 
candidates in AC114, and 7 in Abell 1835 (4 first priority targets and 3
secondary ones; Pell\'o et al. 2004); 2/3 of objects in this sample display
emission lines. The efficiency of our survey nowadays could range between $\sim 30$
and 50\%, with interesting low-$z$ by-products. A large majority of our
high-$z$ candidates still need to be confirmed, either by a redetection of the
faint emission line, or by the non-detection of other lines expected at low-$z$.
 
The results presented here are to be confirmed in different ways. An enlarged
spectroscopic survey is urgently needed to determine the efficiency of our selection
technique. Also, increasing the number of lensing fields with ultra-deep
near-IR photometry is essential to get tighter constraints on the abundance
and physical properties of $z\ge 7$ starburst galaxies.

\acknowledgements{Based on
observations collected at the European Southern Observatory, Chile
(069.A-0508,070.A-0355,073.A-0471), the NASA/ESA Hubble Space
Telescope operated by the Association of Universities for Research in
Astronomy, Inc., and the Canada-France-Hawaii Telescope operated by
the National Research Council of Canada, the French Centre National
de la Recherche Scientifique (CNRS) and the University of Hawaii.
Part of this work was supported by the CNRS and the Swiss National
Foundation.}

\vfill 
\end{document}